\RequirePackage[2020-02-02]{latexrelease}
\documentclass[%
 reprint,
 amsmath,amssymb,
 aps,superscriptaddress,
]{revtex4-2}

\usepackage{graphicx}
\usepackage{dcolumn}
\usepackage{bm}
\usepackage{xcolor}
\usepackage[displaymath, mathlines]{lineno}

\begin{document}

\preprint{APS/123-QED}

\title{Exclusion of ALP Cogenesis Dark Matter in a Mass Window Above 100 $\mu$eV}
\author{Aaron Quiskamp}
\email{aaron.quiskamp@research.uwa.edu.au}
\affiliation{Quantum Technologies and Dark Matter Laboratory, Department of Physics, University of Western Australia, 35 Stirling Highway, Crawley WA 6009, Australia.}
\author{Ben T. McAllister}
\email{ben.mcallister@uwa.edu.au}
\affiliation{Quantum Technologies and Dark Matter Laboratory, Department of Physics, University of Western Australia, 35 Stirling Highway, Crawley WA 6009, Australia.}
\affiliation{ARC Centre of Excellence for Dark Matter Particle Physics,
Swinburne University of Technology, John St, Hawthorn VIC 3122, Australia}
\author{Paul Altin}
\affiliation{ARC Centre of Excellence For Engineered Quantum Systems, The Australian National University, Canberra ACT 2600 Australia}
\author{Eugene N. Ivanov}
\affiliation{Quantum Technologies and Dark Matter Laboratory, Department of Physics, University of Western Australia, 35 Stirling Highway, Crawley WA 6009, Australia.}
\author{Maxim Goryachev}
\affiliation{Quantum Technologies and Dark Matter Laboratory, Department of Physics, University of Western Australia, 35 Stirling Highway, Crawley WA 6009, Australia.}
\author{Michael E. Tobar}
\email{michael.tobar@uwa.edu.au}
\affiliation{Quantum Technologies and Dark Matter Laboratory, Department of Physics, University of Western Australia, 35 Stirling Highway, Crawley WA 6009, Australia.}

\date{\today}

\begin{abstract}
We report the results of Phase 1b of The ORGAN Experiment, a microwave cavity haloscope searching for dark matter axions in the $107.42-111.93~\mu$eV mass range. The search excludes axions with two-photon coupling $g_{a\gamma\gamma}\geq 4\times 10^{-12}\, \textrm{GeV}^{-1}$ with $95\%$ confidence interval, setting the best upper bound to date and with the required sensitivity to exclude the axion-like particle cogenesis model for dark matter in this range. This result was achieved using a tunable rectangular cavity, which mitigated several practical issues that become apparent when conducting high mass axion searches, and was the first such axion search to be conducted with such a cavity. It also represents the most sensitive axion haloscope experiment to date in the $\sim100~\mu$eV mass region.

\end{abstract}

\maketitle

Although direct detection remains elusive, cosmological observations and early Universe simulations allow us to indirectly infer the presence of dark matter, which is thought to constitute $\sim85\%$ of the total matter and $\sim27\%$ of the total energy density in our Universe \cite{planck}. The axion particle is one of the most compelling candidates for dark matter and was originally postulated by Peccei and Quinn as a solution to the strong CP (charge-parity) problem in quantum chromodynamics (QCD) \cite{PQ1977,Weinberg1978,Wilczek1978}. 

Axions can be detected using a kind of experiment called a resonant cavity haloscope, first proposed by Sikivie in 1983 by exploiting their expected coupling to electromagnetism \cite{Sikivie83haloscope,Sikivie1984}. When immersed in a strong DC magnetic field, axions are expected to create detectable photons via the inverse Primakoff effect, with a frequency $\nu_a = m_a c^2/h + \mathcal{O}(10^{-3}c)$, where $m_a$ is the axion rest mass and $\mathcal{O}(10^{-3}c)$ is the expected velocity dispersion that results in an effective axion quality factor, $Q_a$ \cite{isothermal_halo_1990,isothermal_halo_2003}. Tuning the resonant cavity to this frequency enhances the signal by the cavity quality factor, $Q_0$. However, because the axion mass and the axion-photon coupling $g_{a\gamma\gamma}$ are unknown, a wide parameter space spanning several orders of magnitude must be scanned. Although unconstrained by theory, most models favour an axion with a mass above $\mathcal{O}(\mathrm{\mu eV})$ \cite{Preskill1983,graham_2018} and below $\mathcal{O}(\mathrm{meV})$ \cite{Preskill_1982,Keil_1997}. 

For example, the Standard Model Axion Seesaw Higgs portal inflation (SMASH) model is a well motivated description of particle physics that predicts axion dark matter particles to have mass $50\,\mu\mathrm{eV} \leq m_a \leq 200\, \mu\mathrm{eV}$ \cite{SMASH2017,SMASH2019}. The Kim-Shifman-Vainshtein-Zakharov (KSVZ) and Dine-Fischler-Srednicki-Zhitnitsky (DFSZ) models are two of the most popular QCD axion models and are parametrised by the dimensionless coupling constant $g_{\gamma}$, taking values of $-0.97$ and $0.36$ respectively \cite{K79,Zhitnitsky:1980tq,DFS81,SVZ80,Dine1983}. Together with the axion decay constant $f_a$, and the fine structure constant $\alpha$, they determine the axion-photon coupling $g_{a\gamma\gamma} = \alpha g_{\gamma} / \pi f_a$. The axion converted photon power at these couplings is extremely weak $\sim \mathcal{O}(10^{-22}-10^{-23}\,\mathrm{W})$, and only two experiments so far have attained the required sensitivity: ADMX \cite{Du_ADMX,Braine2020,Bartram2021,ADMX2021} and CAPP-12TB \cite{CAPP_DFSZ_2023}. 

The expected axion-photon signal power on resonance extracted by the haloscope is given by \cite{Kim2020}

\begin{equation}
    P_{\mathrm{a\rightarrow\gamma}}=\bigg( g_{a \gamma \gamma}^{2} \frac{\rho_{a}}{m_{a}}\bigg)\bigg(\frac{\beta}{1+\beta} B_{0}^{2} V C Q_{L}\bigg).
    \label{signal_power}
\end{equation}

Here $\rho_a \approx 0.45 \mathrm{GeV/cm^3}$ is the local dark matter density (assumed to be all axions), $\beta$ is the coupling of the ``strongly" coupled antenna port to the haloscope, $Q_L$ is the resonator loaded quality factor, $V$ is the cavity volume and $C$ is the form factor which represents the degree of overlap between the cavity electric field and the externally applied magnetic field, with field strength $B_0$. 

The Oscillating Resonant Group AxioN (ORGAN) Experiment is a microwave cavity haloscope hosted at the University of Western Australia which targets its search for axions in the $62-207~\mu$eV (15-50 GHz) mass region, encompassing the majority of the predicted SMASH range \cite{ORGAN_1a,MCALLISTER201767}. Phase 1a of ORGAN was recently completed, scanning over the mass range between $63.2 -67.1~\mu$eV and setting an upper limit on axion-photon couplings $|g_{a\gamma\gamma}| \geq 3\times 10^{-12}$ in this range \cite{ORGAN_1a}. This search was sufficient to exclude the axion-like particle (ALP) cogenesis model, which is another well motivated model that predicts much stronger axion-photon coupling than QCD axion models \cite{Co2021,Co2020,Co2020b}. 

\begin{figure}[t!]
    \centering
    \includegraphics[width=\linewidth]{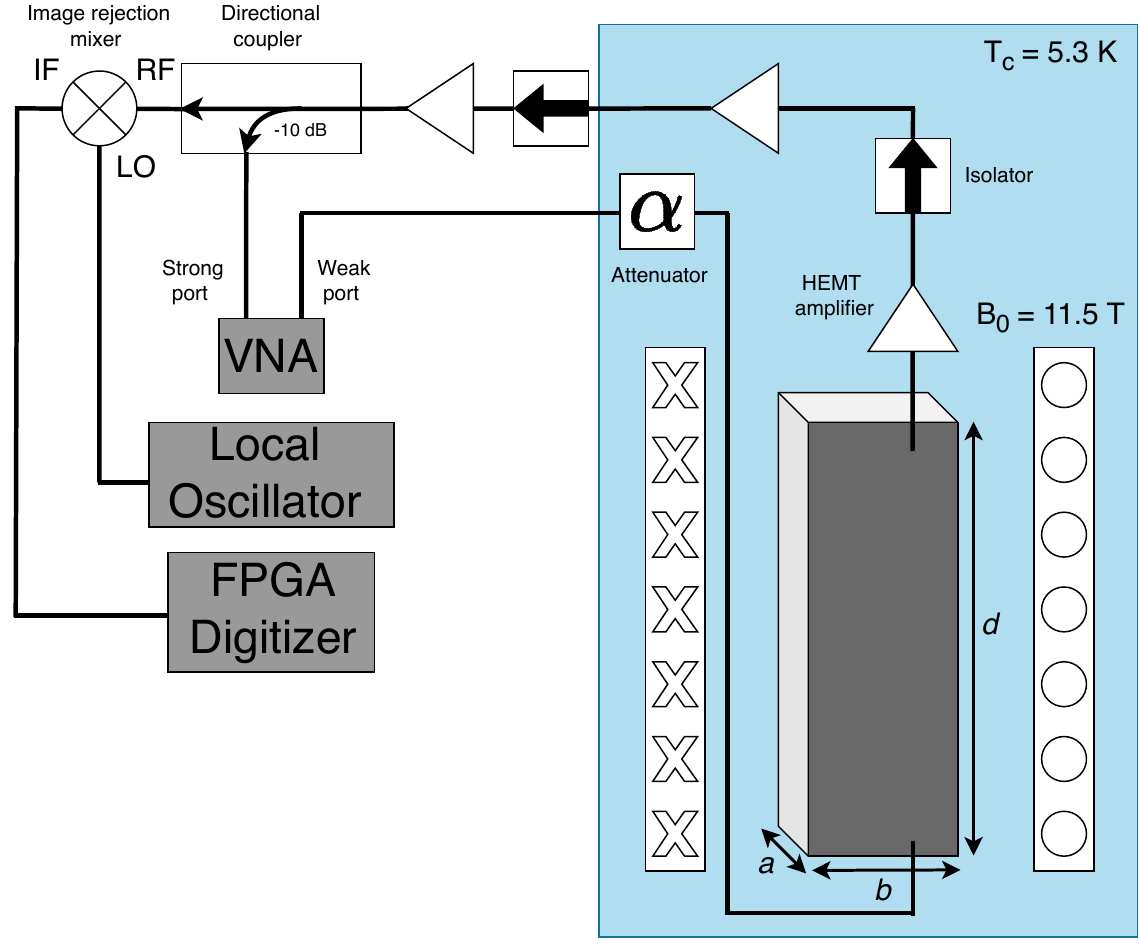}
    \caption{A diagram of the ORGAN Phase 1b experiment. The rectangular cavity at physical temperature $T_C=5.3$ K with dimensions $a, b$ and $d$ as mentioned in the text is placed inside a solenoidal magnet with a field strength of 11.5 T. The tunable wall is shown shaded in dark grey. An expected axion signal would first be amplified through a series of HEMT amplifiers, and later mixed down to an intermediate frequency (IF) using an image rejection mixing stage and local oscillator (LO) for sampling by a field-programmable gate array (FPGA) digitizer. The frequency response of the TM$_{110}$ mode in transmission is measured using a vector network analyzer (VNA).}
    \label{fig:setup}
\end{figure}

In this work, we report the results of the ORGAN Phase 1b experiment (depicted in Fig. \ref{fig:setup}), which achieves ALP cogenesis exclusion over the $107.42-111.93~\mu$eV ($25.97-27.07$ GHz) mass range, whilst operating at a physical temperature of $\sim 5.3$ K and in an 11.5-T magnetic field. First stage amplification is achieved using a low-noise high electron mobility transistor (HEMT) amplifier, which is directly connected to the strongly coupled port of the cavity for negligible loss.

The signal-to-noise ratio (SNR) of a haloscope to a given \cite{Dicke_radiometer}, 
 
\begin{equation}
    \textrm{SNR}=\frac{P_{a\rightarrow\gamma}}{k_B T_\mathrm{sys}} \sqrt{\frac{\tau}{\Delta \nu_{a}}},
    \label{SNR}
\end{equation}

where $k_B$ is Boltzmann's constant, $T_{\mathrm{sys}}$ is the total system noise temperature, $\tau$ is the integration time and $\Delta \nu_a = \nu_c/Q_a \sim \nu_c / 10^{6}$ is the expected axion linewidth given by the isothermal, vitalized halo model for axion dark matter \cite{isothermal_halo_1990,isothermal_halo_2003}. The total system noise temperature is thus a critical parameter in haloscope sensitivity. In Phase 1b, similarly to Phase 1a, we express the total system noise as \cite{amp_noise_absorb}, 

\begin{equation}
	T_{sys}=T_C	\frac{4 \beta}{(1+\beta)^{2}+4\Delta^{2}}+T_A \frac{1+4\Delta^{2}}{(1+	\beta)^{2}+4\Delta^{2}}.
	\label{Tsys}
\end{equation}

Here $T_A$ is the added noise from the first stage amplifier (the contribution to the total added noise from subsequent amplification stages are suppressed by the first stage gain and so can be safely ignored), $T_C$ is the physical cavity temperature, and $\Delta=(\nu-\nu_c)/\Delta\nu_c$ is the detuning from resonance normalised to the cavity linewidth $\Delta\nu_c$.  

To date, the most sensitive searches have primarily utilised cylindrical tuning-rod-based resonators which employ a TM$_{010}$ mode. Such resonators are capable of maximising $B_0^2 V$ in typical, cylindrical solenoidal magnets, and these designs have achieved state-of-the-art sensitivity in the mass range around a few $\mu$eV. However, this type of design is subject to significant challenges, particularly in the move to higher axion masses (and equivalently higher resonant frequencies). Challenges such as tuning rod misalignment and sticking at cryogenic temperatures (reducing $C$ and $Q_0$), radiation leakage as a result of external mechanical tuning (reducing $Q_L$), and poor tuning rod thermalisation (increasing $T_{sys}$) can occur in these resonators, and are common experimental hurdles. These factors together dramatically degrade the achievable scan rate. For these and other reasons, alternate experimental approaches have been proposed - including dielectric and multi-cell haloscopes, and other designs \cite{multi_cell,Quiskamp2020,Supermode2018,MADMAX2017,Orpheus_2022}. 

In ORGAN Phase 1b we opt for a different style of resonator in an attempt to mitigate the challenges above, some of which are expectedly amplified in the high frequency regime owing to the necessarily small volume. A rectangular oxygen-free-high conductivity copper cavity was chosen, where the axion sensitive TM$_{110}$ mode was tuned by moving a single cavity wall perpendicular to the $B^z$ field direction. The cavity dimensions of $a=5.6-6$ mm in the tunable $\vec{x}$ direction, $b=27.1$ mm in the $\vec{y}$ direction and $d=77.3$ mm in the $\vec{z}$ direction were chosen such that no mode crossings occur in the targeted $26-27$ GHz scan range. Although this design suffers some volume loss when compared to an equivalent cylindrical resonator, it can be partially or fully compensated by the increased form factor, with $C_{110}=0.65$. Furthermore, the straightforward frequency tuning mechanism in the form of a moveable wall which is only required to move a fraction of a mm within the cavity reduces complexity, which is critical at cryogenic temperatures. The design also maintains good thermalisation as no weak link to the tuning mechanism is required. Additionally, the absence of a tuning rod avoids the introduction of transverse electromagnetic (TEM) modes, while also reducing longitudinal symmetry breaking and therefore also reduces the strength and number of mode crossings. Further details of this haloscope design, particularly in comparison to the standard tuning-rod haloscope can be found in \cite{McAllister_rectangle}. 

ORGAN Phase 1b acquired data in May 2023 for $\sim 28$ days including rescans, covering all axions masses between $107.42-111.93 \, \mu$eV ($25.97-27.07$ GHz) with sufficient sensitivity to exclude the ALP cogenesis model over the entire range. Again, there were no mode interactions in the range, and thus no forbidden frequencies. The strong port coupling was set statically and characterised in a dedicated cool-down and experimental run prior to data-taking, which when combined with $Q_L$ measurements gave an estimate of the intrinsic quality factor $Q_0$. Using this mapping between $Q_L$, $\nu_c$ and $\beta$, the antenna coupling for the data taking run can be inferred at each cavity position directly from the $\textit{in-situ}$ measurement of $Q_L$. Over the course of the scan, $Q_L$ had a typical value of $\sim 3700$, and $\beta$ was found to lie between $0.19-2.18$ with a mean value of $1.22$. The average deviation in $Q_L$, and the subsequently interpolated $\beta$ values at a given cavity frequency between the two cool-downs, were $4.6\%$ ($\sigma = 2.5\%$) and $8.8\%$ ($\sigma=5.2\%$) respectively. As such, we are able to estimate the coupling to within a tolerable uncertainty without the need to measure it directly $\textit{in-situ}$. 

As discussed, frequency tuning was achieved by moving a single cavity wall in the $\vec{x}$ direction via an Attocube ANPz101 piezoelectric stepper motor mounted externally to the cavity. The tunable wall had 0.1 mm gaps between all four sides of the wall and cavity, allowing consistent and repeatable tuning at cryogenic temperatures.  Because the physical cavity dimensions were changed in order to tune the resonant frequency, so too was the volume, which for the frequency range scanned varied between 11.8 mL and 12.3 mL. 

\begin{figure}[b!]
    \centering
    \includegraphics[width=0.65\linewidth]{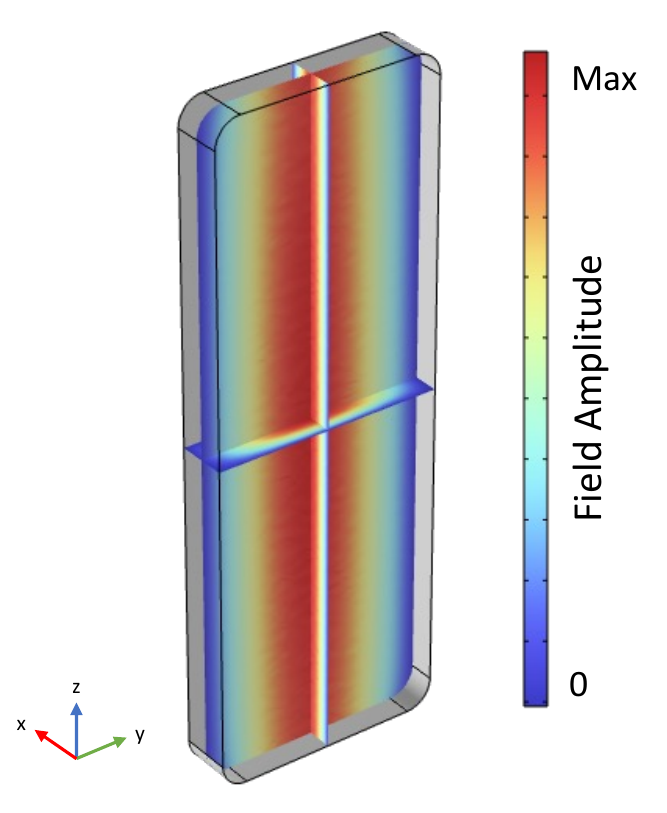}
    \caption{An isometric view of the $z$ component of the TM$_{110}$ electric field. The resonant frequency of the rectangular cavity is tuned via movement of one of the side walls.}
    \label{fig:comsol_model}
\end{figure}

\begin{figure}[t!]
    \centering
    \includegraphics[width=\linewidth]{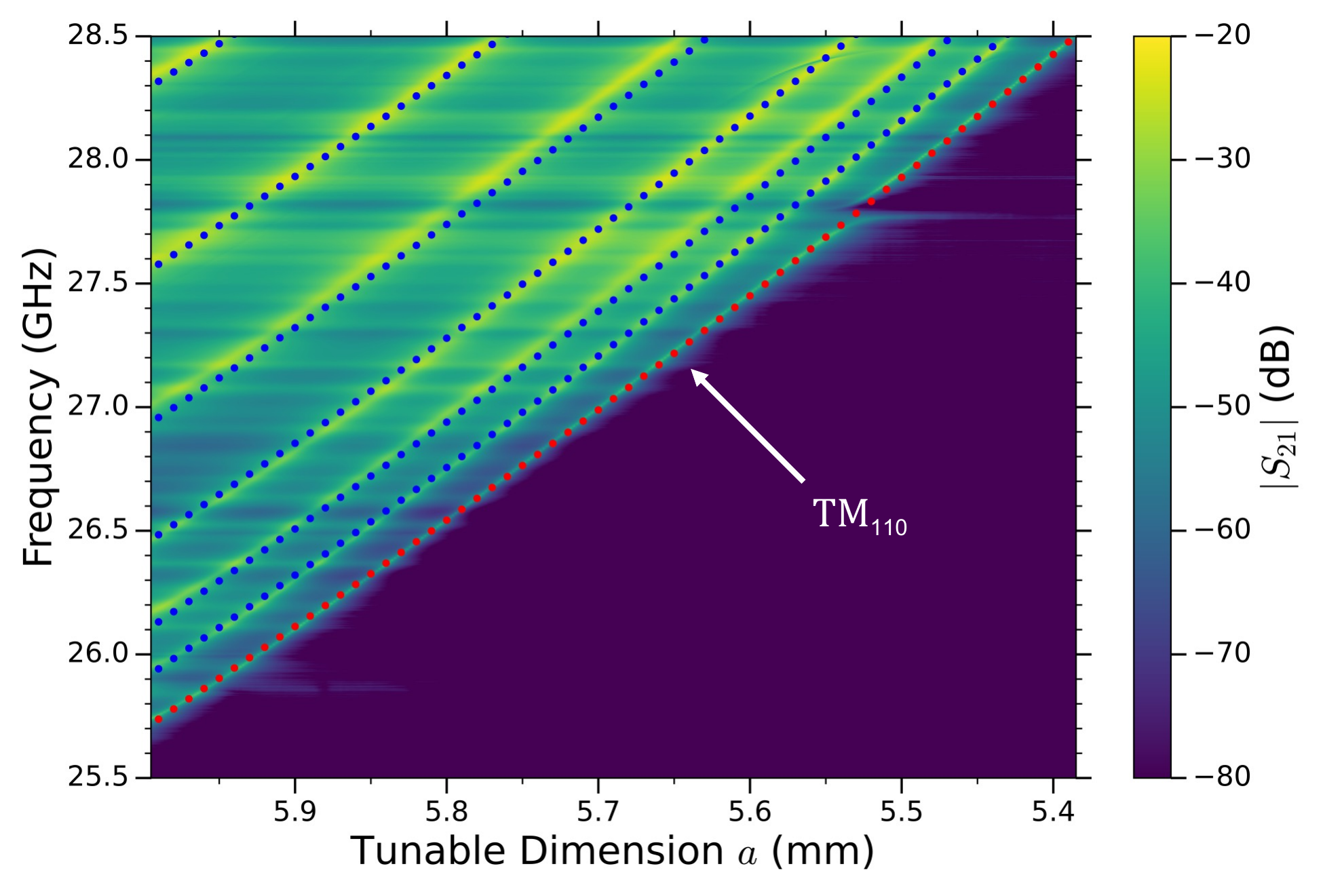}
    \caption{A colour density plot of the transmission coefficient $|S_{21}|$ (dB) as a function of frequency and tunable cavity dimension, $a$. The axion sensitive TM$_{110}$ mode is annotated and tunes over the targeted $26-27$ GHz range with no mode crossings. The broad horizontal lines represent standing waves in the cavity-receiver chain. The dots represent COMSOL modelling of the TM$_{110}$ mode (red) and the first six TM$_{11p}$ modes (blue).}
    \label{fig:modemap}
\end{figure}
 
The form factor $C$ was calculated to be $\sim 0.40$ using Finite Element Method (FEM) modelling in COMSOL Multiphysics. The reduction in $C$ compared to the theoretical maximum value of $\sim 0.65$ for this cavity geometry arises from tilt in the tunable wall. The degree of tilt was quantified by minimising the measured frequency deviations between the TM$_{11p}$ mode family and theoretical modelling in COMSOL, which is displayed in Fig. \ref{fig:modemap}. By considering tilt of the wall in the $\vec{x}$ direction, we find the minimum average deviation between the first six TM$_{11p}$ ($p\geq1$) modes to be $0.087\%$, which occurs when the degree of tilt from vertical is $\sim 0.067^{\circ}$. In principle, this tilt can be mitigated through enhanced mechanical design, for example ultra-low run-out stepper motors, lighter tunable walls and improved assembly tolerances. Such adjustments are planned for such future searches. We believe that the degree of tilt can reduced to less than $\sim 0.02^{\circ}$, corresponding to $C\geq0.58$ with these improvements. However, it should be noted that the degradation in $C$ due to wall tilt is dependent on the wall position since an equivalent degree of tilt has a greater relative effect with decreasing cavity dimensions. 

\begin{figure*}[t!]
    \centering
    \includegraphics[width=\linewidth]{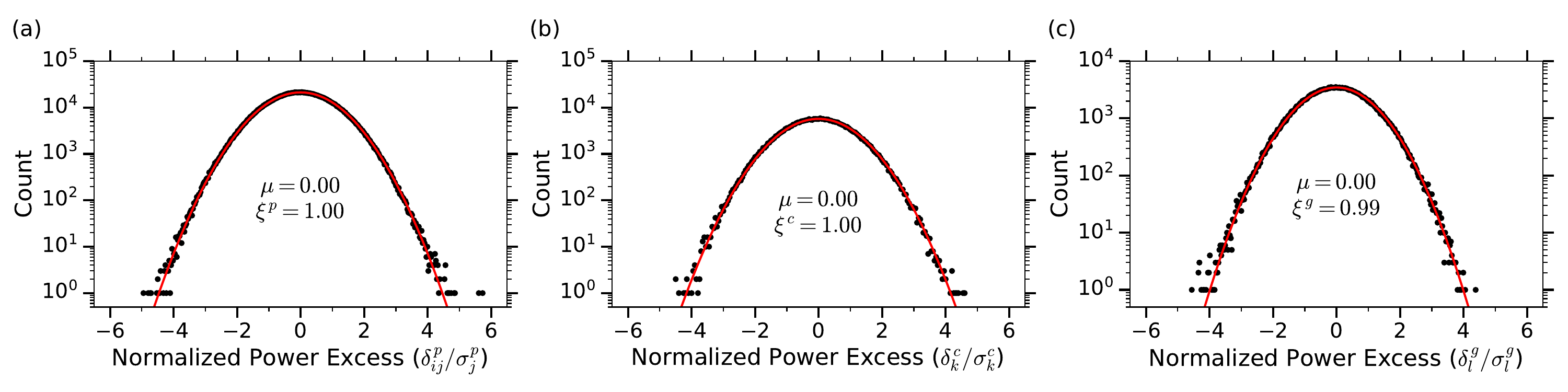}
    \caption{Distributions of the normalised power excess from all frequency bins (black circles) are shown at different stages of the co-adding procedure, with the corresponding Gaussian fits (solid red). (a) Normalised excess power in the $j$th bin of the $i$th SG filtered spectrum $\delta^p_{ij}/\sigma^p_j$ shows the expected Gaussian distribution with $\mu=0$ and  width $\xi^p=1$. (b) The normalised excess power in the $k$th vertically combined bin is given by $\delta^c_k/\sigma^c_k$, and preserves the Gaussian noise with $\mu=0$ and $\xi^c=1$. (c) The distribution of normalised grand spectrum bins $\delta^g_l/\sigma^g_l$ after the horizontal co-addition and optimal filtering deviates from the standard normal Gaussian distribution with a reduced width $\xi^g=0.99$. This results from the negative correlations between adjacent co-added bins that is induced by the SG-filter.}
    \label{fig:histogram}
\end{figure*}

\begin{figure*}[t!]
    \centering
    \includegraphics[width=0.85\linewidth]{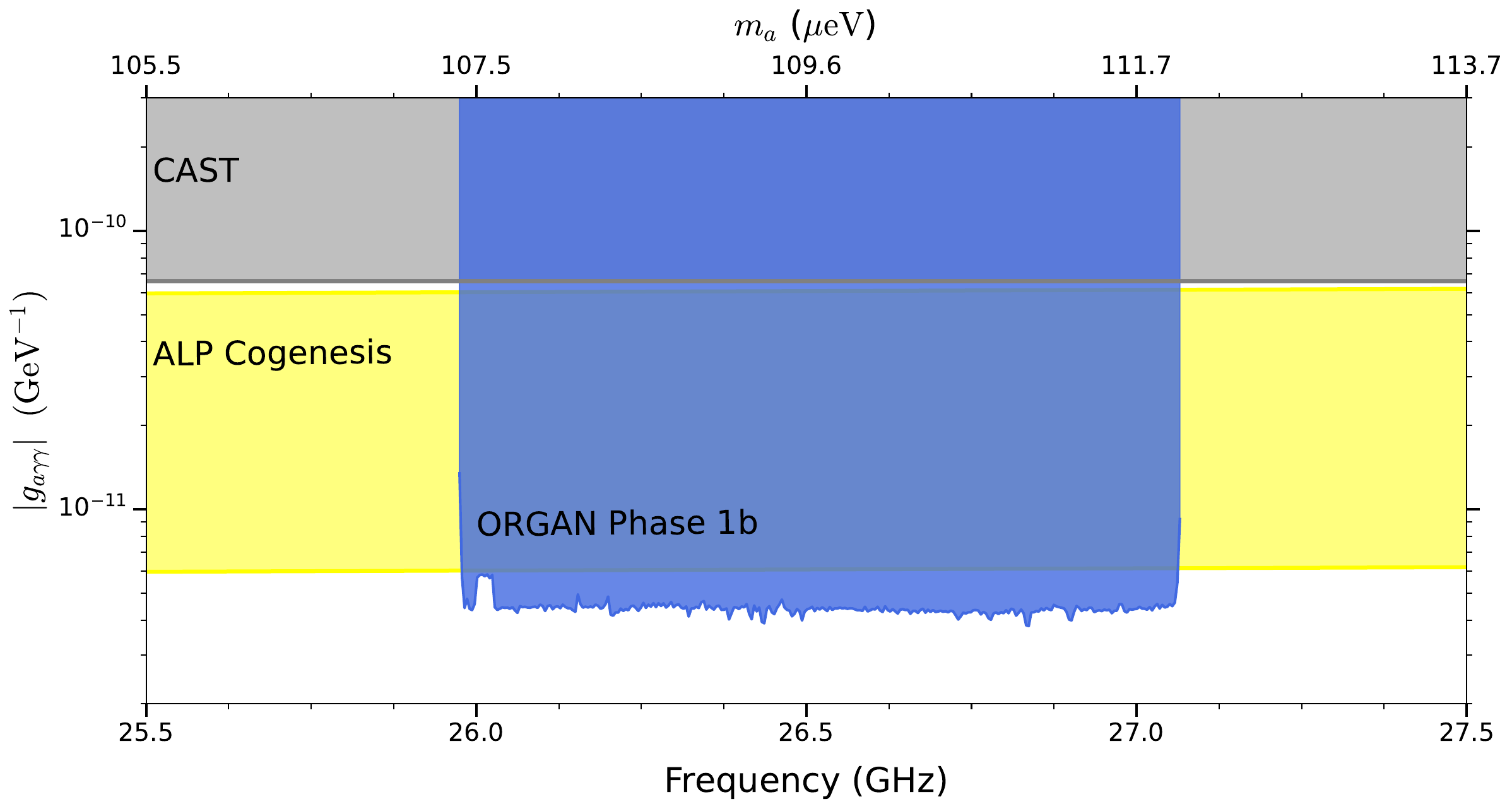}
    \caption{The upper exclusion limits on $g_{a\gamma\gamma}$ from ORGAN phase 1b at a $95\%$ CL is shown shaded in blue. Previous limits set by CAST (grey) \cite{CAST2017} and ALP cogenesis model band (yellow) \cite{Co2021,Co2020,Co2020b} are also shown. The notch at $\sim26$ GHz is due to insufficient integration time in this region.}
    \label{fig:limits}
\end{figure*}

The readout chain, as shown in Fig. 1, consisted of a vector network analyzer (VNA) that measured the frequency response of the TM$_{110}$ mode in transmission in order to find $Q_L,\, \nu_c$ and $\Delta\nu_c$ at each tuning step. After the first amplification stage, isolators were used between stages to prevent back-action noise. The down-mixing stage achieved image rejection of the noisy, amplifier-only sideband by driving two identical mixers with the same local oscillator (LO) source except $90^{\circ}$ out of phase. The intermediate frequency (IF) outputs from both mixers were recombined using a $90^{\circ}$ hybrid coupler and sampled by a 250 MS/s digitizer (NI-5761R) with real time data processing done on a field-programmable gate array (FPGA; Xilinx Kintex-7, NI-7935R). A zero dead-time, hybrid superheterodyne fast Fourier transform spectrum analyzer was implemented on the FPGA, generating a 6,553 point, 25 MHz-wide spectrum centered at 45 MHz, with a bin width of $\Delta\nu_b \approx 3815$ Hz. The analysis window was cropped to be $9.91$ MHz wide to avoid harmonics of the $10$ MHz reference clock signal at 40 and 50 MHz. 

The integration time at each cavity frequency varied between between 38 and 211 minutes and  was adjusted based on $V$, $Q_L$ and $\beta$ to maintain sensitivity better than the ALP cogenesis model bands. The data analysis procedure follows the method outlined by The Haloscope At Yale Sensitive To Axion CDM (HAYSTAC) Experiment \cite{Brubaker2017b}, and is the same method used in the ORGAN Phase 1a analysis \cite{ORGAN_1a}. The slowly varying baselines of the raw power spectra are removed with a Savitsky-Golay (SG) filter, while preserving spectral features on the scale of $\Delta\nu_a \sim 26-27$ kHz, which equates to $\sim 7\Delta\nu_b$ for this search. The SG filter had a polynomial degree of 5 and a 401 point window size. However, the SG filter is known to attenuate potential axions signals as a result of its imperfect stop band attenuation \cite{Brubaker2017b}. Through simulation we find the attenuation of an axion signal due to SG-filtering to be $\eta_{SG} = 0.95$, which reduces our exclusion power by the same factor. After normalising via the SG filter and subtracting 1 so that the spectra have $\mu=0$, we obtain a set of dimensionless, normalised excess power spectra and in the absence of axion conversion, each bin is a Gaussian random variable with $\mu=0$ and standard deviation $\sigma^p=1/\sqrt{\Delta\nu_b \tau}$. The processed spectra are subsequently rescaled according to their expected axion sensitivity such that overlapping spectra can be vertically combined using a maximum likelihood weighted sum which maximises SNR. However, as mentioned the axion line shape is expected to follow a Maxwell-Boltzmann distribution and so bins must be co-added horizontally to match this width and shape. We enhance our sensitivity to axion detection by optimally filtering sets of 7 consecutive bins according to this expected line shape, resulting in a normalised ``grand power spectrum" which is shown as a histogram in Fig. \ref{fig:histogram}. A rescan threshold of $4~\sigma$ was set, which 5 spectra exceeded. 

The potential axion candidates were rescanned with sufficient integration time so that after subjecting the rescanned spectra to analysis, none were found to still exceed the $4\,\sigma$ threshold. A $95\%$ confidence level (CL) is chosen and corresponds to an SNR target of $5.645 \sigma$, from which an upper bound on axion-photon coupling is set, assuming axions make up $100\%$ of the local dark matter density of 0.45 GeV/$\mathrm{cm}^3$. As shown in Fig. \ref{fig:limits}, ORGAN Phase 1b surpasses the limits set by The CERN Axion Solar Telescope (CAST) \cite{CAST2017} by over an order of magnitude and excludes ALP cogenesis over the targeted mass range of $107.42-111.93 \,\mu$eV ($25.97-27.07$ GHz). The ``notch" at 26 GHz is due to insufficient integration time early in the data taking run, but the sensitivity is still sufficient to exclude ALP cogenesis in this region. We calculate the fractional uncertainty on axion-photon coupling in the same way as ORGAN Phase 1a \cite{ORGAN_1a}, which takes into account uncertainties in the parameters $\beta \, (8.8\%)$, $Q_L \,(3\%)$, $C \, (5\%)$, $V\,(1\%)$ and $T_A \,(10\%)$. This results in a total uncertainty of $\delta g_{a\gamma\gamma} \approx \, 6.1\%$. 

We have reported on Phase 1b of The ORGAN Experiment, a high mass axion haloscope hosted at The University of Western Australia. Phase 1b excluded ALP cogenesis dark matter in the mass range of $107.42-111.93 \,\mu$eV. With the completion of Phase 1b, the first phase of ORGAN is now complete, and to date has set the most sensitive limits on axion-photon coupling above 10.4 GHz ($43\, \mu$eV) \cite{QUAX_43uev}, excluding ALP cogenesis over the searched ranges. Phase 1b utilised a novel type of resonator, and has demonstrated the feasibility of high mass axion searches despite the unfavourable sensitivity scaling. Future phases will leverage detector and readout upgrades such as new cavity designs, GHz single photon counting, superconducting cavities and a larger magnet to probe deeper into the parameter space between $15-50$ GHz. 

\begin{acknowledgments}
This work is supported by Australian Research Council Grants CE17010009 and CE200100008.
\end{acknowledgments}

\end{document}